\documentclass[twocolumn,showpacs,preprintnumbers,floatfix]{revtex4}
\usepackage{graphicx}

\def\la{\langle }
\def\ra{\rangle }
\def\be{\begin{equation}}
\def\ee{\end{equation}}
\def\bea{\begin{eqnarray}}
\def\eea{\end{eqnarray}}

\begin{document}

\preprint{astro-ph/0312124}

\title{Can the Local Supercluster explain the low CMB multipoles?}

\author{L. Raul Abramo}
\email{abramo@fma.if.usp.br}
\affiliation{Instituto de F\'{\i}sica, 
Universidade de S\~ao Paulo \\
CP 66318, 05315-970 S\~ao Paulo, Brazil}

\author{Laerte Sodr\'e Jr.}
\email{laerte@astro.iag.usp.br}
\affiliation{Instituto Astron\^omico e Geof\'{\i}sico, 
Universidade de S\~ao Paulo \\
CP 3386, 01060-970 S\~ao Paulo, Brazil}

\date{\today} 

\begin{abstract} 
We show that the thermal Sunyaev-Zeldovich effect caused by hot 
electrons in the Local Supercluster (LSC) can explain the 
abnormal quadrupole and octopole of the cosmic microwave 
background (CMB) that were measured by WMAP and COBE.
The distortion needed to account for the low observed quadrupole is a 
spot in the direction
of the LSC with a temperature decrease of order 
$\Delta T \approx - 7 \, \mu$K for $\nu \sim $ 20 --- 90 Ghz photons.
The temperature and density of the hot gas which can generate such an 
effect are consistent with observations of the X-ray background.
If this hypothetic foreground
is subtracted from the WMAP data, we find that the 
amplitude of the quadrupole ($\ell=2$) is 
substantially increased, and that the ``planarity'' of both 
the quadrupole and the octopole ($\ell=3$) are weakened.
For smaller scales the effect decays and, at least in our
simplified model, it does not affect the angular power 
spectrum at $\ell > 10$.
Moreover, since the Sunyaev-Zeldovich effect 
increases the temperature
of photons with frequencies above 218 GHz, observations sensitive in that
range (such as PLANCK's HFI) will be able to confirm whether the LSC 
indeed affects the CMB.
\end{abstract}


\maketitle

\section*{Introduction}

The cosmic microwave background (CMB) anisotropies have now been measured
with exquisite accuracy by WMAP \cite{WMAP1}. Such a barrage of new data
seldom brings only confirmation of known theories and mechanisms, 
and WMAP is no exception: early reionization \cite{WMAP1},
lack of higher correlations \cite{Correl}
and a curious supression of power at the largest observable
scales \cite{OTZH,Efstathiou} 
are some of the most intriguing questions that have 
been raised by the WMAP data.
In this letter we focus on the problem of the 
CMB multipoles corresponding to the largest scales, and 
show that at least these anomalies can be
explained by ordinary physics.

\vskip 0.2cm

The CMB temperature anisotropies on very large scales were first 
measured by COBE \cite{COBE}. WMAP \cite{WMAP1} confirmed those 
observations and showed moreover a nearly flat curve of 
the angular power spectrum $C_\ell$ at large scales (spherical
harmonic indices $\ell < 100$)
and a pattern of acoustic oscillations at smaller scales ($\ell > 150$).
This is consistent with the inflationary picture of a nearly-scale 
invariant spectrum of adiabatic density perturbations.

However, the data is not entirely devoid of its quirks: there are
a few ``sticky'' points in the observed angular power spectrum, 
in particular those around $\ell=200$, $\ell=40$ and $\ell=20$, all with
statistically significant deviations from the expected (smooth) curve. 
In addition to those points, WMAP confirmed the COBE observation that
showed that the quadrupole ($\ell=2$) appears to be 
supressed by a factor $\sim$  80\%
with respect to nearby multipoles.
Furthermore, an analysis of the components $a_{\ell m}$
for the quadrupole and octopole reveals that both have an unusual 
degree of symmetry (``planarity'') \cite{OTZH}.

The actual relevance of the deviant data points seems to be still under 
debate: different estimates for the chance that the low 
value of the quadrupole can be explained by a purely statistical
fluctuation vary, from 0.15\% \cite{WMAP2} to 
5\% \cite{OTZH,Efstathiou} to 30\% \cite{Wagg}
--- incidentally, these are much less significant factors than obtained 
for the ``sticky'' points at $\ell$=40 and $\ell$=200.
This lack of power at large scales has motivated many ingenious
explanations, such as compact topologies \cite{Luminet}, a broken or
supressed spectrum at large scales \cite{Broken}
and oscillations superimposed on the primordial spectrum of density
fluctuations \cite{Jerome}.

However, when the low value of the quadrupole is combined with
the unusual symmetry of the quadrupole and octopole ($\ell=3$),
and with the alignment of the directions defined by these two
multipoles, the overall chance of such a statistical fluctuation
falls to 0.004\% \cite{OTZH}. As noted by
de Oliveira-Costa {\it et al.} \cite{OTZH}, the directions
preferred by the quadrupole and the octopole 
point roughly towards the Virgo cluster --- 
which is, of course, the direction of the dipole as well.
Add to this indications that the polarization of radio and optical 
sources also have a tendency to point in that same direction
\cite{RalstonJain}, and the reported
differences in the CMB maps
between the northern/western galactic hemispheres (where lie
Virgo and most of the LSC) and the southern/eastern hemispheres 
\cite{Correl}, and the string of coincidences
becomes rather too long to ignore.

We propose here that the explanation for the properties of the
quadrupole and octopole is a chance alignment of a hot spot
of the primordial temperature fluctuations with the region 
of the sky occupied by the local supercluster (LSC) --- 
which is centered roughly around Virgo.
The thermal Sunyaev-Zeldovich effect (SZe) due to hot electrons 
in the intra-supercluster (ISC) medium causes, for the range of 
frequencies observed by WMAP and COBE, an apparent decrease in the 
temperature of the CMB photons in the direction of the LSC, 
with an amplitude which we have estimated, using a simple model, 
as being of order $|\Delta \hat{T}_{\ell=2}|_{rms} \approx 7$ $\mu$K
for the quadrupole, and similar (but smaller) values for $\ell>2$.

This means that the primary anisotropies of the CMB
could actually be interfering with the SZe caused
by the LSC, so that the observed
low multipoles of the CMB are significantly distorted
with respect to their true (primordial) values. This distortion would
supress the quadrupole and introduce a preferred direction in the 
components of the low multipoles --- 
which would, of course, point towards Virgo.

\section*{Sunyaev-Zeldovich effect in the LSC?}

The SZe is caused by the inverse Compton scattering
of CMB photons by hot electrons in the intra-cluster medium \cite{SZ}. 
It is a {\it nonthermal}, frequency-dependent effect:
the upscattering causes an incident blackbody spectrum of photons to 
become distorted in such a way that the resulting 
higher abundance of high-energy photons 
is compensated by a shortage of low-energy photons. The
frequency at which photons are neither depleted nor overproduced
is $\nu_0 = 218$ GHz \cite{RevSZ} (COBE DMR and WMAP work in the 
range 20 -- 90 GHz). Therefore, for frequencies below $\nu_0$
the effect is a reduction in the temperature of the photons, and
for frequencies above that the effect is the opposite.
This means that measurements
over a range of frequencies around $\nu_0$ (such as PLANCK's LFI and HFI
\cite{PLANCK}) can pick up the signal of the SZe and
distinguish it from the primary anisotropies.

For low-frequency photons, the SZe is given by:
\be
\label{dT-y}
\frac{\Delta \hat{T} (\theta , \phi)}{T_{0}} = - 2 y(\theta,\phi) \; ,
\ee
where $T_{0}=2.726$ K is the CMB temperature and 
$y$ is the comptonization parameter in the direction
$(\theta,\phi)$. The comptonization parameter $y$ measures an
optical depth for the CMB photons created by the hot electrons, 
and its value is given by the product of the Thomson cross-section 
$\sigma_T = 6.65 \times 10^{-25}$ cm$^2$
by the temperature-averaged density of photons along the line of sight
\cite{RevSZ}:
\be
\label{y}
y = \int \sigma_T \frac{k T_e}{m_e c^2} n_e dl \; ,
\ee
where $T_e$ is the electron temperature, $m_e$ is the electron mass
and $dl=dl(\theta,\phi)$ is the line-of-sight distance element along
the direction $(\theta,\phi)$.

The SZe has been observed over the past few years in many clusters, 
but its weak strength means that it could only be
detected in the central parts of clusters, where column densities
of hot gas are sufficiently high \cite{RevSZ,SZObs}.
It is evident that some amount of SZ will take place also
in the LSC, but the question is, how much? The answer depends on the 
gas density in the ISC medium, its temperature distribution, 
the morphology of the LSC and our position inside it.

The morphology of the LSC is relatively well known \cite{LSCmorph}:
it is a flattened collection of groups and clouds of galaxies 
centered at the Virgo Cluster, which contains $\sim$20\% of 
its bright galaxies.
The Local Group is dynamically linked to the LSC, and lies
$\sim$15 Mpc from Virgo, at the border of the LSC.
Notice that the LSC itself is not a virialized structure,
hence the gas in its midst is not necessarily in equilibrium.

We are interested in an analytic approach at this point,
hence we will make a radical simplification by
approximating the shape of the LSC by an oblate spheroid of maximal 
radius 20 Mpc with approximate axial ratios 6:3:1 \cite{LSCmorph}.
Therefore, our simple model assumes that the LSC is a collection
of objects (clouds, groups and the Virgo cluster) which are
distributed smoothly across the spheroid.
The Sun stands at the margin of 
the spheroid (which looks like a flattened pumpkin), 
approximately 15 Mpc away from Virgo (see Fig. 1.)
\begin{figure}
\includegraphics[width=8cm]{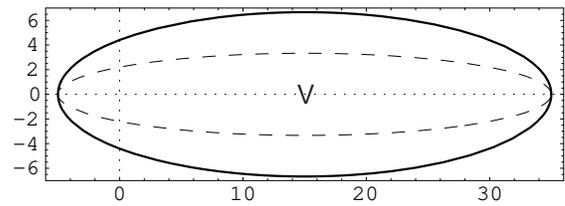}
\caption{\label{fig:1} Assumed shape of the LSC (solid and dotted 
ellipsoids denote different cross-sections of the spheroid.) 
Distances are in Mpc. The Sun is located
at the origin and Virgo (V) is at the center of the LSC.}
\end{figure}

Much less known than the shape of the LSC, however, 
are the density and temperature distribution of the 
hot gas of the ISC medium. 
%The main direct evidence for 
%hot electrons in {\it clusters} comes from emission of X-rays due to 
%bremstrallung: correlating X-ray with microwave data,
%one can try to map the density of hot gas in clusters. 
Unfortunately, X-ray and microwave observations have not yet 
reached the level of sensitivity required
to detect directly the very smooth, diffuse columns of hot gas in 
the outer regions of clusters.
It seems, however, obvious that there must be a great amount
of ionized gas in the ISC medium, among other reasons because 
the absence of the Gunn-Peterson effect means that most of the 
hydrogen that we know must exist is not in neutral form. The gas
is thought to have been shock-heated at the time of galaxy formation, and
now it is probably distributed in many phases, including filaments and 
a more homogeneous component \cite{CO,Kravtsov,Sembach}.
Phillips, Ostriker and Cen \cite{POC} have constrained the amount 
of gas in filaments 
%(the ``warm-hot intergalactic medium'') 
using numerical simulations and 
the X-ray background, and argued that this ``warm-hot'' 
($kT \approx$ 100 eV -- 10 keV) gas can account for only 5--15 \% 
of the ``missing baryons''. It is therefore quite possible that much 
of this gas is in the ISC medium.
The question is then, how hot is this ionized gas, and how is it 
distributed?

Hogan was the first to propose 
that superclusters (and the LSC) could 
impact the CMB anisotropies through the thermal and kinetic
Sunyaev-Zeldovich effects \cite{Hogan}.
Molnar and Birkinshaw used HEAO 1 A2 \cite{HEAO} and COBE DMR data to 
analyze the Shapley supercluster and found no evidence of hot 
($>10^7$K) gas in the ISC medium \cite{MB}.
%Their result agreed with Rephaeli \cite{Rephaeli}, and with
%Banday and Gorski \cite{Banday}, who did not find
%evidence for a halo of hot electrons in the local group of galaxies.
Boughn \cite{Boughn}, on the other hand,
used the HEAO 1 A2 X-ray map and a simple ``pillbox'' model
of nearly constant electron density in the LSC to argue that the
SZe could be as high as 
$|\delta T| \sim (17 \pm 5)$ $\mu$K --- although he assumed
a gas temperature in the high end of the range $10^5$--$10^8$ K.
Kneissl {\it et al.} \cite{Kneissl} 
did study the correlation of COBE DMR and ROSAT X-ray
data away from the galactic plane, but it is not clear that the X-ray 
data has enough sensitivity to detect the diffuse hot gas of the LSC, 
and in any case the authors analyzed a region which 
misses a large chunk of the LSC.
% --- they limited their analysis to the slice 
%$70^o < l < 250$, while the LSC center
%is located around $l=250^o$. 

Much work has been done to study the impact of the SZe from
{\it distant} clusters on the CMB \cite{Diaferio,MB2,Uros,SR}.
It has been found that the
largest contribution to the angular power spectrum from the SZe
comes from the most massive clusters 
($M \sim 10^{15} h^{-1} M_\odot$), at scales $\ell \sim 3000$, with 
amplitudes
$\ell(\ell+1)C_\ell/2 \pi \approx 10 - 100 \, \mu$K$^2$.
%Given that the low-$\ell$ plateau of the CMB spectrum lies at 
%$\ell(\ell+1)C_\ell/2 \pi \approx 800 \mu$K$^2$, one could suppose
%that the SZe from a nearby cluster would be too weak to 
%distort the observed temperature anisotropies at lower $\ell$'s.
%These estimates are for the
%contribution from distant clusters {\it averaged over the whole sky}.
%Since we live inside one such massive supercluster, with 
%many groups of galaxies and at least one very massive cluster,
%the SZe from those sources can have a much stronger signal.

%We have nothing to add to this important issue at this point: only
%future multi-band observations of the CMB and better X-ray
%measurements can decide whether or not the LSC is a
%relevant source of secondary anisotropies.

\section*{Angular power spectrum of the LSC SZe}

The overall number of free electrons in the LSC can be computed
given its mass: $N_e = M_{LSC} f_g / \mu_e m_p$, where $M_{LSC}$ 
is the LSC mass, $f_g$ is the gas fraction, 
$\mu_e$ is the molecular weight per electron and 
$m_p$ is the proton mass. 
We may assume that the mass of the LSC is
$\sim 7 \times 10^{15}$ M$_\odot$ \cite{SPT}. 
Assuming that the Hydrogen is fully ionized
and that the helium mass fraction is
$Y=0.24$, then $\mu_e = 1/(1-Y/2) \simeq 1.14$. 
The gas fraction is not very well known, but X-ray observations of 
clusters indicate that $f_g \approx 0.06 h^{-3/2}$ \cite{EttoriFabian99}. 
Using $h=0.7$ we get finally that the total number of electrons in the
LSC should be of order $N_e \sim 7 \times 10^{71}$.
We assume that the volume of the LSC is $V_{LSC} = 4 \pi/3 \times A B C$,
where $A=20$ Mpc, $B=6.7$ Mpc and $C=3.3$ Mpc are 
the principal semi-axes of the spheroid, the 
average density of electrons in the LSC is 
$n_e = N_e/V_{LSC} \sim 1.4 \times 10^{-5}$ cm$^{-3}$.

A convenient approximation is to assume a constant electron density across
the LSC. If the gas has an {\it average} temperature
of 2 keV then $\la k T_e \ra / m_e c^2 \simeq 0.004$, and
with a line-of-sight distance of $30 \, {\rm Mpc}$ we obtain that
the comptonization parameter is of order
$ \Delta y \approx \sigma_T \la k T_e \ra/(m_e c^2)
\, \times n_e \times 30 {\rm Mpc} \approx 3.5 \times 10^{-6} $.

The amplitude that is needed to explain the quadrupole is of
order $\Delta y \approx 10^{-5}$. Therefore, if that is the case then
either the gas is hotter than 2 keV, or the density of electrons
is higher than $10^{-5}$ cm$^{-3}$, or both.

The main constraint on the density and temperature of the ISC medium comes
from the X-ray background. A compilation of observations [20] gives a
background flux for energies $h \nu \sim 1$ keV of approximately $10^{-25}$
erg s$^{-1}$ cm$^{-2}$ sr$^{-1}$ Hz$^{-1}$. 
On the other hand, the expected
flux at this energy due to thermal bremsstrahlung emission in the center of 
a uniform sphere 
of radius 30 Mpc with $n_e = 1.4 \times 10^{-5}$  cm$^{-3}$ and $T_e = 2$ keV 
is $\sim 2 \times 10^{-26}$ erg s$^{-1}$ cm$^{-2}$ sr$^{-1}$ Hz$^{-1}$.
Since the X-ray flux is proportional to the square of the electron density,
if the gas temperature is indeed 2 keV, the upper bound for the
electronic density is of order $n_e \approx 5 \times 10^{-5}$ cm$^{-3}$.

The comptonization parameter can be exactly computed from Eq. (\ref{y})
for our ``pumpkin model''. The assumed ionized gas distribution is 
uniform inside the oblate spheroid defined by 
$(6x)^2+(3y)^2+z^2=A^2$, and zero outside it.
The angular power spectrum for the SZe of the LSC 
is given in Fig. 2.
The amplitude of the SZe quadrupole is:
\bea
\label{quadrupole}
\Delta \hat{T}_2^2 &\equiv& \frac{6}{2 \pi} \hat{C}_2 
\approx 50 \, \alpha^2 \, \mu{\rm K}^2 \; ,\\
\nonumber
\alpha &=&  \frac{n_e}{5 \times 10^{-5} {\rm cm}^{-3}} \times
\frac{\la k T_e \ra}{ 2 \, {\rm keV}} \; .
\eea
This level of temperature distortion agrees with the COBE FIRAS 
limit on deviations
from the blackbody spectrum on large angular scales \cite{FIRAS}.

Compare the results in Fig. 2  
with the WMAP data points for $\ell < 20$, for which
$\ell(\ell+1)C_\ell/2\pi \approx 800$ $\mu$K$^2$ \cite{WMAP1}. 
The SZe quadrupole has almost the same order of magnitude 
as the expected primary CMB quadrupole.
The SZe octopole ($\ell=3$) appears to be very small, but
the multipoles $\ell>2$ are more sensitive to the assumed symmetry.
Nevertheless, the fall-off with $\ell$ 
is expected, given our assumption of a homogeneous
gas distribution. Hence, even if the SZe from
the LSC contributes at the largest scales, that effect becomes
irrelevant as $\ell$ grows, as long as the ISC gas is diffuse enough.
Another interesting result of our calculations is the fact that
the amplitude of the $m=0$ components seem to be 
higher than the amplitudes of the $m \neq 0$ components.
However, a more precise statement 
concerning the decomposition of the SZe from the LSC
into components $\hat{a}_{\ell m}$
is evidently not possible until we consider a more
realistic approach to the gas density and temperature
distributions in the LSC.

\begin{figure}
\includegraphics[width=8cm]{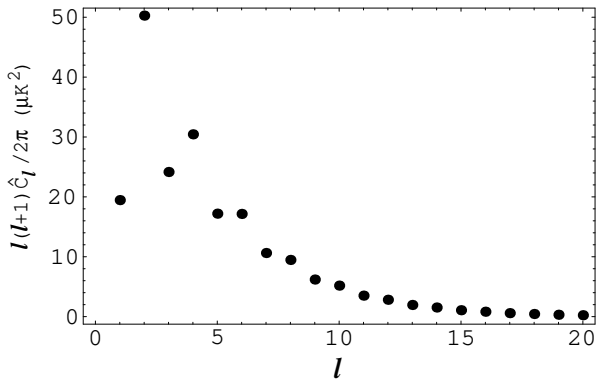}
\caption{\label{fig:2} Angular power spectrum
of the LSC SZe (with $\alpha=1$.)}
\end{figure}

%A geometric factor which may increase the amplitude of the SZe
%from the nearby distribution of matter 
%is the existence of the
%local void roughly opposite to the position of the Virgo cluster
%\cite{Naka}. In any case, the detection of a SZe due to the
%LSC will bring strong contraints on the physical conditions of gas
%in the ISC medium.

The interference between the SZe from the LSC and
the primordial CMB can be estimated, if one rotates the
axes of the CMB maps so that they coincide with ours.
This task is facilitated because de Oliveira-Costa {\it et al.}
\cite{OTZH} have computed the
components of the quadrupole and octopole of the 
temperature fluctuations observed by WMAP, $a_{\ell m}^{O}$,
in the rotated frame whose $z$-axis points in the 
direction of Virgo. If the $x$- and $y$-axis coincided as well,
we could subtract the computed
$\hat{a}_{\ell m}$ directly from the observed $a_{\ell m}^{O}$, 
to obtain the primordial multipoles.
But since we do not know the precise angles between the two reference
systems, the $m\neq 0$ components have unknown phases 
$\exp{[i m \Delta \phi]}$ between them.
For the quadrupole of the SZe we get:
\be
\label{quad_2}
\hat{a}_{20} \approx - 15 \, \alpha \, \mu{\rm K} \quad , \quad \hat{a}_{21} 
\approx 0 \quad , \quad |\hat{a}_{22}| \approx 4 \, \alpha \, \mu{\rm K} \; .
\ee
Assuming $\alpha=1$ and
combining these components with those given in \cite{OTZH},
we obtain an estimate for the amplitude of the primordial quadrupole:
\be
\Delta T_2^2 = \frac{6}{2\pi} C_2 \approx (230 - 370) 
\, \mu{\rm K}^2 \; ,
\ee
where the range corresponds to an unknown 
phase relating the $m=\pm 2$ components. If either the temperature
or the density of electrons are higher than our fiducial values
such that $\alpha=4$, we would 
get that $\Delta T^2_2 \approx 340-640$ $\mu$K$^2$.
The amplitude of the octopole also grows after subtraction of the 
SZe foreground, but 
by a smaller factor which appears to be more model-dependent.

After subtraction of the SZe foreground, 
the levels of symmetry of the corrected quadrupole and octopole 
seem to be lower than those of the observed
quadrupole and octopole. It should be possible to detect
the same effect in the higher $\ell$ components as well, but
the lower amplitude of the SZe and the larger number of 
components, which grow as $2\ell+1$, 
can make such a distinction harder to establish.
In any case, the precise way in which the SZe from the LSC
breaks into harmonic components 
is evidently quite sensitive to the morphology of the LSC and its 
spatial orientation.
A more careful analysis is being carried out, combining the
CMB maps with the observed LSC morphology, which will compute
in detail the corrected CMB angular power spectrum.

%\vskip 0.5cm

\section*{Conclusions}

We have argued that the SZe from the LSC
can affect the low multipoles of the anisotropies of the 
CMB. The temperature and density of
the hot LSC gas which causes the SZe obey the
observational constraints on the X-ray background.
%We used a very simple model for 
%the shape of the hot gas distribution in the LSC to 
%estimate that effect. 
After we subtract this hypothetical foreground, we obtain 
a greatly increased quadrupole and less symmetric components 
for the quadrupole and octopole. 
We can interpret the result for the quadrupole
as meaning that there is a
large-scale hot spot in the primordial CMB which roughly coincides
with the position of the LSC. The probability that 
such an alignment happens by chance
is of order 10-20\%.

If the LSC indeed affects the temperature anisotropies through
the thermal SZe, then it is conceivable that the {\it kinetic} 
SZe (caused by the anisotropic motion of gas),
which has typical amplitudes one order 
of magnitude weaker than the thermal SZe, could be
important as well \cite{Hogan}. Since the kinetic SZe polarizes
the CMB photons, this might have interesting implications for
the WMAP detection of polarization and reionization.

We should note that the evidence for hot gas in the ISC
medium that can cause such effects is still weak. However, near-future
X-ray and millimiter-band observations in the region of the
North galactic plane will easily decide this issue. In particular,
PLANCK's planned observations over a wide range of frequencies 
(30 -- 900 GHz) will be able to clearly pick any SZe signal \cite{PLANCK}.
If the SZe is indeed observed in the LSC, it would have many 
interesting implications for the physical properties of hot gas in 
the ISC medium.

\vskip 0.2cm

\noindent
We would like to thank G. Holder for pointing errors in our earlier
drafts. We also thank R. Rosenfeld for many fruitful
conversations, and S. Boughn, M. Coutinho, 
G. Hinshaw, M. Tegmark and I. Waga for useful comments.
This work was supported by FAPESP and CNPq.

%\section{Appendix}

%\vspace{.1cm}


\begin{thebibliography}{99}

\bibitem{WMAP1} C. Bennett {\it et al.}, {\it Astrophys. J. Suppl.} 
{\bf 148}: 1 (2003).
%%CITATION = ASTRO-PH 0302207;%%

\bibitem{Correl} 
H. Eriksen {\it et al.}, astro-ph/0307507.
%%CITATION = ASTRO-PH 0307507;%%

\bibitem{OTZH} A. de Oliveira-Costa, M. Tegmark, M. Zaldarriaga and A. 
Hamilton, astro-ph/0307282.
%%CITATION = ASTRO-PH 0307282;%%

\bibitem{Efstathiou} G. Efstathiou, 
astro-ph/0310207.
%%CITATION = ASTRO-PH 0310207;%%

\bibitem{Wagg} E. Gazta\~nagam {\it et al.},
{\it MNRAS} {\bf 346}: 47 (2003).
%%CITATION = ASTRO-PH 0304178;%%

\bibitem{COBE} K. Gorski {\it et al.}, {\it Astrophys. J.} {\bf 464}: L11 
(1996);
%%CITATION = ASTRO-PH 9601063;%%
E. Wright {\it et al}, {\it Astrophys. J.} {\bf 464}: L21 (1996).
%%CITATION = ASTRO-PH 9601059;%%

\bibitem{WMAP2}
D. Spergel {\it et al.}, {\it Astrophys. J. Suppl.} {\bf 148}: 175 (2003).
%%CITATION = ASTRO-PH 0302209;%%

\bibitem{Luminet}
J.-P. Luminet, J. Weeks, A. Riazuelo, R. Lehoucq and J.-P. Uzan,
{\it Nature} {\bf 425}: 593 (2003).
%%CITATION = ASTRO-PH 0310253;%%

\bibitem{Broken} C. Contaldi, M. Peloso, L. Kofman and A. Linde,
{\it JCAP} {\bf 0307}: 002 (2003);
%%CITATION = ASTRO-PH 0303636;%%
J. Cline, P. Crotty and J. Lesgourgues,
{\it JCAP} {\bf 0309}: 010 (2003);
%%CITATION = ASTRO-PH 0304558;%%
B. Feng and X. Zhang, {\it Phys. Lett. } {\bf B570}: 145 (2003);
%%CITATION = ASTRO-PH 0310206;%%
%%CITATION = ASTRO-PH 0305020;%%
S. Tsujikawa, R. Maartens and R. Brandenberger,
{\it Phys. Lett.} {\bf B574}: 141 (2003).
%%CITATION = ASTRO-PH 0308169;%%

\bibitem{Jerome}
J. Martin and C. Ringeval, astro-ph/0310382.
%%CITATION = ASTRO-PH 0310382;%%

\bibitem{RalstonJain} J. Ralston and P. Jain, astro-ph/0311430.
%%CITATION = ASTRO-PH 0311430;%%

\bibitem{SZ}
R. Sunyaev and Ya. Zeldovich, {\it Ap{\&}SS} {\bf 7}: 3 (1970);
{\it Comments Astrophys. Space Phys.} {\bf 4}: 173 (1972).

\bibitem{RevSZ}
J. Carlstrom, G. Holder and E. Reese, {\it Annu. Rev. Astron. Astrophys.}
{\bf 40}: 643 (2002); M. Birkinshaw, astro-ph/0307177.
%%CITATION = ASTRO-PH 0307177;%%
%%CITATION = ASTRO-PH 0208192;%%

\bibitem{PLANCK} http://astro.estec.esa.nl/SA-general/Projects/Planck/

\bibitem{SZObs}
L. Grego {\it et al.}, {\it Astrophys. J.} {\bf 552}: 2 (2001); 
B. Mason, S. Myers and A. Readhead, {\it Astrophys. J.} {\bf 555}: L11 (2001);
M. de Petris {\it et al.}, {\it Astrophys. J.} {\bf 574}: L119 (2002).
%%CITATION = ASTRO-PH 0012067;%%
%%CITATION = ASTRO-PH 0101169;%%
%%CITATION = ASTRO-PH 0203303;%%

\bibitem{LSCmorph} R. Tully, {\it Astrophys. J.} {\bf 257}: 389 (1982);
G. Giuricin, C. Marinoni, L. Ceriani and A. Pisani,
{\it Astrophys. J.} {\bf 543}: 178 (2000);
C. Marinoni, G. Giuricin, L. Ceriani, and A. Pisani, in
``Cosmic Flows Workshop'', ASP Conference Series, Vol. 201, Ed. 
S. Courteau and J. Willick (2000), astro-ph/9909444.
%%CITATION = ASTRO-PH 9909255;%%
%%CITATION = ASTRO-PH 9909444;%%

\bibitem{CO} R. Cen and J. Ostriker,
{\it Astrophys. J.} {\bf 514}: 1 (1999); {\it ibidem}, {\bf 517}: 31 (1999).

\bibitem{Kravtsov} A. Kravtsov, A. Klypin and Y. Hoffman,
{\it Astrophys. J.}{\bf 571}: 563 (2002). 
%%%CITATION = ASTRO-PH 0109077;%%

\bibitem{Sembach} K. Sembach, astro-ph/0311089.
%%CITATION = ASTRO-PH 0311089;%%

\bibitem{POC} L. Phillips, J. Ostriker and R. Cen, {\it Astrophys. J.}
{\bf 554}: L9 (2001).

\bibitem{Hogan} C. Hogan, {\it Astrophys. J.} {\bf 398}: L77 (1992).

\bibitem{HEAO} S. Pravdo {\it et al.}, {\it Astrophys. J.} {\bf 234}: 
1 (1979).

\bibitem{MB} S. Molnar and M. Birkinshaw, {\it Astrophys. J.} {\bf 497}:1 
(1998).

\bibitem{Boughn} S. Boughn, {\it Astrophys. J.} {\bf 526}: 14 (1999).

\bibitem{Kneissl} R. Kneissl, R. Egger, G. Hasinger, A. Soltan and J.
Tr\"umper, {\it Astron. Astrophys.} {\bf 320}: 685 (1997).
%%CITATION = ASTRO-PH 9610160;%%

\bibitem{Diaferio}
A. Diaferio, A. Nusser, N. Yoshida and R. Sunyaev,
{\it MNRAS} {\bf 338}: 433 (2003); A. Diaferio, R. Sunyaev
and A. Nusser, {\it Astrophys. J.} {\bf 533}: L71 (2000).
%%CITATION = ASTRO-PH 9912117;%%
%%CITATION = ASTRO-PH 0207420;%%

\bibitem{MB2} S. Molnar and M. Birkinshaw, {\it Astrophys. J.} {\bf 537}: 
542 (2000).
%%CITATION = ASTRO-PH 0002271;%%

\bibitem{Uros} E. Komatsu and U. Seljak, {\it MNRAS} {\bf 336}: 1256 (2002);
{\it MNRAS} {\bf 327}: 1353 (2001).
%%CITATION = ASTRO-PH 0106151;%%
%%CITATION = ASTRO-PH 0205468;%%

\bibitem{SR} S. Sadeh and Y. Rephaeli, {\it New Astron.} {\bf 9}: 159 (2004).

\bibitem{SPT} E. Shaya, P. J. Peebles and R. Tully, {\it Astrophys.
J.} {\bf 454}: 15 (1995).

\bibitem{FIRAS} D. Fixsen, G. Hinshaw, C. Bennett and J. Mather, 
{\it Astrophys. J.} {\bf 486}: 623 (1997); D. Fixsen, 
{\it Astrophys. J.} {\bf 594}: L67 (2003).

%\bibitem{Naka} K. Nakanishi {\it et al.}, 
%{\it Astrophys. J. Suppl} {\bf 112}: 245 (1997).

\bibitem{EttoriFabian99} 
S. Ettori and A. Fabian, {\it MNRAS} {\bf 305}: 834 (1999).
%%CITATION = ASTRO-PH 9901304;%%


%\bibitem{NOIS} L. R. Abramo, R. Rosenfeld and L. Sodr\'e Jr, to appear.

\end{thebibliography}
\end{document}